# Single-layer spin-orbit-torque magnetization switching due to spin Berry curvature generated by minute spontaneous atomic displacement in a Weyl oxide

*Hiroto Horiuchi,[*] Yasufumi Araki, Yuki K. Wakabayashi, Jun'ichi Ieda, Michihiko Yamanouchi, Shingo Kaneta-Takada, Yoshitaka Taniyasu, Hideki Yamamoto, Yoshiharu Krockenberger, Masaaki Tanaka,[*] and Shinobu Ohya[*]*

H. Horiuchi, S. Kaneta-Takada, M. Tanaka, S. Ohya
Department of Electrical Engineering and Information Systems
The University of Tokyo
7-3-1 Hongo, Bunkyo-ku, Tokyo 113-8656, Japan
E-mail: h-horiuchi22@g.ecc.u-tokyo.ac.jp, masaaki@ee.t.u-tokyo.ac.jp, ohya@cryst.t.u-tokyo.ac.jp

Y. Araki, J. Ieda
Advanced Science Research Center
Japan Atomic Energy Agency
2-4 Shirakata, Tokai-mura, Naka-gun, Ibaraki 319-1195, Japan
Email: araki.yasufumi@jaea.go.jp

Y. K. Wakabayashi, Y. Taniyasu, H. Yamamoto, Y. Krockenberger
NTT Basic Research Laboratories
NTT Corporation
3-1 Morinosato Wakamiya, Atsugi-shi, Kanagawa 243-0198, Japan
Email: yuuki.wakabayashi@ntt.com

M. Yamanouchi
Division of Electronics for Informatics, Graduate School of Information Science and Technology
Hokkaido University
Kita 14 Nishi 9, Sapporo-shi, Hokkaido 060-0814, Japan

M. Tanaka, S. Ohya
Center for Spintronics Research Network (CSRN)
The University of Tokyo
7-3-1 Hongo, Bunkyo-ku, Tokyo 113-8656, Japan

M. Tanaka, S. Ohya
Institute for Nano Quantum Information Electronics, The University of Tokyo, 4-6-1 Komaba, Meguro-ku, Tokyo 153-8505, Japan

**Spin Berry curvature characterizes the band topology as the spin counterpart of Berry curvature and is crucial in generating novel spintronics functionalities. By breaking the crystalline inversion symmetry, the spin Berry curvature is expected to be significantly enhanced; this enhancement will increase the intrinsic spin Hall effect in ferromagnetic materials and, thus, the spin–orbit torques (SOTs). However, this intriguing approach has not been applied to devices; generally, the extrinsic spin Hall effect in ferromagnet/heavy-metal *bilayer* is used for SOT magnetization switching. Here, SOT-induced partial magnetization switching is demonstrated in a *single* layer of a single-crystalline Weyl oxide $SrRuO_3$ (SRO) with a small current density of ~$3.1 \times 10^6$ A cm$^{-2}$. Detailed analysis of the crystal structure in the seemingly perfect periodic lattice of the SRO film reveals barely discernible oxygen octahedral rotations with angles of ~5° near the interface with a substrate. Tight-binding calculations indicate that a large spin Hall conductivity is induced around small gaps generated at band crossings by the synergy of inherent spin–orbit coupling and band inversion due to the rotations, causing magnetization reversal. Our results indicate that a minute atomic displacement in single-crystal films can induce strong intrinsic SOTs that are useful for spin-orbitronics devices.**

1. **Introduction**

Current-induced spin–orbit torque (SOT) magnetization switching has great promise for attaining high-performance spintronics devices, such as magnetoresistive random access memory[1], nano oscillators[2], and logic devices[3]. Generally, ferromagnet (FM)/heavy metal (HM) bilayer systems are used for SOT magnetization switching. In those systems, a spin current is generated from an in-plane current in the HM layer through the spin Hall effect (SHE), exerting torques on the magnetization in the FM layer. However, when HMs with strong spin–orbit coupling (SOC) are used, the critical current density required for switching (~$10^7$ A cm$^{-2}$) is too high for practical applications[4–6]. This drawback partially arises from spin scattering at the FM/HM interface, which suppresses the SOTs. Moreover, the required switching current density scales with the FM layer thickness, while reducing the thickness increases the bit error rates. Recent breakthroughs have demonstrated magnetization switching in FM *single* layers[7–23], providing a promising alternative by simplifying the layer structure of the SOT devices. However, these systems require intentional breaking of the inversion symmetry (IS) to generate SOTs, such as introducing composition gradients in the normal direction of the film[10,12–15,17–23]. This complexity poses significant challenges in maintaining film quality and often leads to undesired spin scattering/relaxation within the films[24,25]. Consequently, a novel SOT-induced switching mechanism that operates at lower critical current densities than conventional bilayer systems while utilizing a simplified, single FM layer structure is urgently needed.

To address these challenges, a promising approach lies in controlling the generation and distribution of the intrinsic SHE within an FM layer while maintaining its crystal quality. The spin Berry curvature[26] characterizes the band topology as the spin counterpart of the Berry curvature and determines the intrinsic SHE. Strong spin Berry curvature is predicted to appear at band crossings with band inversion in materials with strong SOC[27], generating sizable SOTs[28]. The oxide Weyl ferromagnet SrRuO$_3$ (SRO) has a strong SOC and emerges as a particularly intriguing candidate for this approach. It exhibits both ferromagnetism and linear band crossings in the bulk state with spatial IS[29–31]. If this spatial IS can be broken while maintaining the crystal quality of SRO, a great enhancement of SOTs is expected, which will enable efficient *single-layer* magnetization switching. Previous studies have reported that intentionally introduced magnetic domain walls break the IS of SRO, inducing efficient domain-wall motion[32,33]. Here, we demonstrate current-induced SOT partial magnetization switching in an epitaxial *single* layer of perpendicularly magnetized oxide Weyl ferromagnet SRO grown on SrTiO$_3$ (STO) (001) (**Figure** 1a). We obtain a small critical current density of ~ $3.1\times10^6$ A cm$^{-2}$ for switching the magnetization states, one order of magnitude smaller than that required for conventional FM/HM bilayer systems[4–6] and other single-layer systems with composition gradients[12,14,17,19,23]. Notably, ~8% of the magnetization of the 26 nm-thick SRO film is stably reversed by an in-plane current. Our analysis using high-angle annular dark field scanning transmission electron microscopy (HAADF-STEM) reveals that the SRO film is a seemingly defect-free single crystal; however, when we closely examine the lattice structure via annular bright-field scanning transmission electron microscopy (ABF-STEM), we find spontaneous oxygen octahedral rotations with the displacement of oxygen atoms of ~ 0.01 nm near the SRO/STO interface. Our theoretical tight-binding calculations reveal that the oxygen octahedral rotations cause band inversion around the small gaps generated at band crossings due to the broken sublattice symmetry. Consequently, the synergy between the originally existing SOC and the newly introduced band inversion generates strong spin Berry curvature, resulting in significant intrinsic SOTs. Our results indicate that SOC in materials accompanied by only a minute displacement of light-element atoms in single crystals can induce strong SOTs, which are beneficial for spin-orbitronics applications. This finding provides a new guiding principle, which utilizes very local breaking of the crystalline symmetry, for designing materials with substantial SOTs.

## 2. Results and Discussion

### 2.1. Sample Preparation and Characterization

We grew an epitaxial SRO film with a thickness of 26 nm on an STO (001) substrate using a machine learning-assisted molecular beam epitaxy (MBE) system[34]. We made a crossbar device using photolithography and Ar ion milling, followed by sputtering of the Ag electrodes as heat sinks (see Figure 1a and Section 4). For the studied SRO film, the residual resistivity ratio (RRR) [35] is defined as the ratio of the resistance at a temperature $T$ of 300 K to that at a $T$ of 3.7 K and is ~15.5. The longitudinal resistivity ($\rho_{xx}$) shows a kink at $T \sim 150$ K, which corresponds to the Curie temperature ($T_C$) (Figure 1b). The SRO film has perpendicular magnetic anisotropy (PMA), as shown by the rectangular hysteresis of the anomalous Hall effect (AHE) in Figure 1c, where the height of the loop is 0.51 Ω at $T$ = 90 K. The Hall resistance $R_H$ is *negatively* proportional to the perpendicular component of magnetization in the temperature range from $T$ = 3.7 K to 120 K (see Supporting Text 1 and Figure S3, Supporting Information).

### 2.2. *Single-layer* SOT magnetization switching in SRO

We performed current-induced SOT magnetization switching measurements in the following steps. Before the measurements, we applied a large external magnetic field of 1 T along the –$z$ (// $[00\bar{1}]$) or +$z$ (// $[001]$) direction (in pseudo-cubic notation) to align the spontaneous magnetization in those directions and then decreased the magnetic field to zero. Due to the negative AHE coefficient, the initial state with the magnetization orientation along the –$z$ direction corresponds to point A in **Figure** 2c,d. The initial state of the +$z$ magnetization direction corresponds to point B in Figure 2e,f. We applied a weak external magnetic field $H_x$ along the $x$ direction to ensure deterministic magnetization switching[4,6,9] (Figure 2b). Under continuous application of $H_x$, a short-pulsed writing current $I_w$, whose current density is $J$, with a pulse width of 0.1 ms, was applied in the $x$ direction (red squares in Figure 2a). Then, following an interval $t_{int}$ of 0.2 s with no current flow, we applied a pulsed reading current $I_r$ of 1.0 mA with a pulse width of 0.1 s (blue squares in Figure 2a) in the $x$ direction. By measuring the $R_H$ under the application of $I_r$, we detected the magnetization state of the SRO film.

An important finding in our study is the appearance of *single-layer* SOT-induced switching of the magnetization state (Figure 2c–f). As shown in Figure 2c,d, when $J$ increases in the positive direction from the initial point A in process 1, $R_H$ undergoes a sudden jump to an $R_H$ value of ~0 Ω at a current density $J$ of ~5 ×10$^6$ A cm$^{-2}$. The SRO films grown on STO substrates often exhibit a stripe-like pattern of ferromagnetic domains with alternating ±$z$ magnetization orientations[36,37], reflecting the atomic steps of the TiO$_2$-terminated STO (001) surface. When these multidomain structures dominate in the film, the $R_H$ approaches ~0 Ω due to magnetization cancellation. We propose that process 1 likely induces a similar multidomain structure, potentially arising from magnetization instability during current flow[38]. Subsequent current reversals (processes 2 and 3) lead to a hysteresis loop in $R_H$ (Figure 2c, d). Similar behavior is observed starting from the initial state B (Figure 2e,f). Thereafter, the $R_H$ follows the same hysteresis loop when repeating processes 1, 2, and 3 (see Figure S4, Supporting Information). The slight shift in the center of the hysteresis loops from an $R_H$ value of 0 depends on $H_x$ and is caused by the small deviation of $H_x$ from the in-plane direction (approximately 6°; see Supporting Text 2 and Figure S5, Supporting Information). The key feature of our result is the polarity change of the hysteresis loops depending on the sign of $H_x$; this feature is a hallmark of deterministic SOT magnetization switching[6,9]. The loop height is always ~ 0.039 Ω thus, ~8% (= 0.039 Ω/0.51 Ω) of the total magnetization of the SRO film is stably reversed by the SOTs induced by the in-plane current. As described later, this partial magnetization switching can be attributed to the octahedral rotations near the SRO/STO interface, leading to magnetization reversal primarily near this interfacial region (insets in Figure 2c,d).

$\Delta R_H$ is defined as the $R_H$ relative to the center of the hysteresis loops during processes 2 and 3; as shown in **Figure** 3a, when $H_x$ increases, the hysteresis loop height initially increases but then decreases with increasing $\mu_0 H_x$ ($\geq$ +40 mT); this phenomenon is caused by the magnetization tilting in the $H_x$ direction and is typical for SOT switching. In Figure 3b, we observe the same counterclockwise $R_H$–$J$ loops for $\mu_0 H_x$ = +10 mT up to 120 K. The SOT magnetization switching loop disappears at $T$ above a $T_C$ of ~150 K. As shown in Figure 3c, $J_c$ decreases with increasing temperature; this result is attributed to the reduced saturation magnetization and weakened magnetic anisotropy (see Figure S2, Supporting Information). The smallest critical switching current density obtained in this study is ~3.1×10$^6$ A cm$^{-2}$ at $T$ = 120 K.

## 2.3. Broken inversion symmetry due to minute displacement of the oxygen atoms

In general, for single-layer SOT magnetization switching, we need to break the IS. In our heterostructure, however, no inversion asymmetry appears inside our high-quality SRO film, as shown in the HAADF-STEM image in **Figure** 4a. Hence, to clarify the cause of the observed partial magnetization switching, we need a more precise analysis of the local crystal structure, which is not discernible in Figure 4a. For this purpose, we utilized ABF-STEM to detect the position of the oxygen atoms (Figure 4c). We find that the oxygen atoms are slightly shifted alternately in the ±$y$ direction, especially near the STO interface; these results indicate that RuO$_6$ octahedral rotations around the $x$-axis occur (see arrows in the inset of Figure 4c). The octahedral rotation of oxygen is sensitive to the epitaxial strain[39,40], oxygen vacancies, and interfacial coupling with octahedra of substrates[41]. The rotation magnitudes differ depending on the $z$ position within the film. Here, using the bond angle $\theta$ of Ru–O–Ru (Figure 4b), we define the oxygen octahedral rotation angle $\alpha$ around the $x$-axis as $\alpha = (180° − \theta)/2$. We find that $\alpha$ sharply increases to ~5° near the SRO/STO interface (Figure 4d). As discussed later, magnetization switching is considered to occur in the 10–14th unit-cell layers counted from the SRO/STO interface just above the peak of $\alpha$ (= 9th layer).

### 2.4. Spin Berry curvature generated by the oxygen octahedral rotations

To understand the influence of the octahedral rotations, we theoretically calculated the spin Berry curvature and spin Hall conductivity (SHC) in SRO with the tilted $RuO_6$ octahedra. The octahedral crystal field splits the Ru $4d$ bands into high-energy $e_g$ and low-energy $t_{2g}$ states. In SRO, electrons exist only in the $t_{2g}$ band. Hence, we constructed a tight-binding model[29,42], with six bases of three $t_{2g}$ orbitals with up and down spins (see Section 4). We considered the band structure of SRO near the SRO/STO interface, where half of the $t_{2g}$ band is filled because of the charge transfer of one electron from Ru to Ti near the SRO/STO interface[43]. Here, $Ru^{4+}$ in SRO has four electrons in the $d$ orbital, whereas $Ti^{4+}$ in STO has none; these configurations are energetically favorable for the transfer of an electron from SRO to STO. This model can successfully reproduce the low-energy band structure, including the Weyl point structure, and the spontaneous magnetization of SRO[29]. In addition, neighboring octahedrons located in the same $xz$ plane rotate in opposite directions at the same angle. Hence, we consider a unit cell consisting of four sublattices 1–4, as shown in **Figure** 5a. Since the volume of the unit cell quadruples, the Brillouin zone is folded into a quarter of that for a single octahedron unit cell, by which band crossings appear at highly symmetric points (see Supporting Text 3 and Figure S6, Supporting Information). In general, the oxygen octahedrons also rotate around the $z$-axis in addition to the $x$-axis; however, the $z$-axis rotation does not significantly change the calculation results (see Figure S7, Supporting Information). Thus, we consider rotation only around the $x$-axis hereafter.

To understand the spin Hall conductivity, we calculated the spin Berry curvatures when $\alpha = 0°$ and $\alpha = 5°$ (Figure 5c,d). We define $E$ and $t$ as the electron energy and electron hopping amplitude of the $\pi$ bonding ($t_\pi^{ij}$) between the nearest neighbor Ru sites, respectively (see Experimental Section). Using the $n$-th eigenvalue $\epsilon_{nk}$ and eigenstate $|u_{nk}\rangle$ of the tight-binding Hamiltonian $\mathcal{H}_k$ at each wave vector $k$, we can calculate the spin Berry curvature $\Omega_{nk}^{y,s_y}$, defined as follows:

$$\Omega_{nk}^{y,s_y} = -2 \sum_{m(\neq n)} \text{Im} \frac{\langle u_{nk}|j_{zk}^{s_y}|u_{mk}\rangle \langle u_{mk}|v_k^x|u_{nk}\rangle}{(\epsilon_{nk} - \epsilon_{mk})^2}, \quad (1)$$

where $v_k^{g=x,y,z}$ is the velocity operator expressed as $\partial \mathcal{H}_k/\partial k_g$ and the $j_{zk}^{s_y}$ is spin–current operator defined as $\frac{1}{2}\{s_y, v_k^z\}$. Here, $\{s_y, v_k^z\}$ is an anticommutator, where $s_y$ is the spin operator in the $y$ direction and $v_k^z$ is the velocity operator in the $z$ direction. As shown in Figure 5c, $E = 0$ is defined as the Fermi level ($E_F$) position corresponding to the half-filled state of the $t_{2g}$ band (see Section 4). At each band crossing obtained when $\alpha = 0°$ near the $E_F$, a small gap opens when $\alpha = 5°$ (especially near the X point, we can see significant changes). Simultaneously, around the opening gaps, band inversion appears, leading to a sharp variation in the wave function in $k$-space. As a synergetic effect with these band modulations, the inherently existing SOC generates *hot spots* of the spin Berry curvature $\Omega_{nk}^{y,s_y}$ (Figure 5c). Accordingly, $\Omega_{nk}^{y,s_y}$ is dramatically increased around these gaps.

To understand the spin Berry curvature distribution around the $E_F$, we examine the Fermi contours in the $k$ plane of $Z'U'R'T'$ (Figure 5b) and the total spin Berry curvature $\Omega_k^{y,s_y} = \sum_n f(\epsilon_{nk}) \Omega_{nk}^{y,s_y}$, where $f(\epsilon_{nk})$ is the Fermi distribution defined as $f(E) = [e^{E/T} + 1]^{-1}$. Here, we take the limit $T = 0$ K. With increasing $\alpha$, we observe an increase in $\Omega_k^{y,s_y}$ around the Fermi contours (see Supporting Text 3 and Figure S8, Supporting Information). Based on a more detailed analysis of the bands, the spin-up and spin-down states are hybridized near the small gaps (see Supporting Text 3 and Figure S9, Supporting Information). These gaps cannot be observed in the absence of octahedral rotations, where the band crossings are protected by sublattice symmetry. Once the sublattice symmetry is broken by the octahedral rotations, the hybridization of the opposite spin bands generates *hot spots* of the spin Berry curvature, where the direction of spins sharply varies in **k**-space, as the synergetic effect with the SOC.

We derive the intrinsic SHC $\sigma_{zx}^{s_y}$, which is defined as the spin-current density with the $y$-component spin flowing in the $+z$ direction divided by the electric field applied in the $x$ direction. By using the Kubo formula, $\sigma_{zx}^{s_y}$ is defined as the integral of the spin Berry curvature over the entire Brillouin zone:

$$\sigma_{zx}^{s_y} = e \sum_n \int_{BZ} \frac{d^3k}{(2\pi)^3} f(\epsilon_{n\mathbf{k}}) \Omega_{n\mathbf{k}}^{y,s_y}. \tag{2}$$

$\sigma_{zx}^{s_y}$ has a sharp peak near $E_F$ ($E/t = 0$ in Figure 5d). With increasing $\alpha$, the magnitude of the peak of $\sigma_{zx}^{s_y}$ becomes larger. From our experimental results, the maximum value of $\sigma_{zx}^{s_y}$ is estimated to be ~9.5×10$^5$ ($\hbar/2e$) $\Omega^{-1}$ m$^{-1}$ at 90 K by using the relations of $\sigma_{zx}^{s_y} = (\hbar/2e)\sigma_{xx}\theta_{SH}$ and $\theta_{SH} = 2eM_s t_{FM} H_c / \hbar J_c$ [44], where $\sigma_{xx}$, $\theta_{SH}$, $M_s$, $t_{FM}$, and $J_c$ are the longitudinal conductivity, spin Hall angle, saturated magnetization, thickness of the FM region where magnetization switching occurs, i.e., ~8% of the 26 nm-thick SRO film (dark red region in Figure 5e), and coercive field, respectively (see Supporting Text 4, Supporting Information).

## 2.5. Mechanism of the unique switching behavior

Figure 5e shows the predicted mechanism of the observed partial SOT magnetization switching. As shown in Figure 2d,f, the magnetization is switched from upward to downward for $J > 0$ when $\mu_0 H_x > 0$, indicating that the damping-like torque by the $-y$-polarized spin is generated when $J > 0$[6]. Based on the theoretical analysis shown above, a large spin current is generated from around the 9th layer from the SRO/STO interface, where the largest octahedral rotation occurs. Due to the positive sign of $\sigma_{zx}^{s_y}$ (see Figure 5d), the spin current diffusing upward has $-y$-direction spin polarization, exerting a large SOT on the magnetization in the 10–14th layers counted from the SRO/STO interface. This estimation is consistent with the region volume estimated from the SOT magnetization switching experiments shown in Figure 2c–f (~8% of the 26-nm-thick SRO film). Similar partial magnetization switching has also been discussed in a ferromagnet with an intentionally controlled composition gradient[23]. The region closer to the SRO/STO interface shown in dark gray in Figure 5e is considered a dead layer of ~3 nm (~8 MLs) with low conductivity, less magnetization, and magnetization instability; this dead layer does not contribute to deterministic switching, as discussed in ref. [45], where magnetization switching does not occur. Consequently, we obtain partial SOT-induced magnetization switching. Based on this result, even a minute spontaneous displacement of oxygen atoms, as small as ~0.01 nm, in our film triggers dramatic band modulation and leads to a substantial intrinsic SHE capable of *single*-layer switching of the magnetization states of ferromagnets.

## 3. Summary

We experimentally demonstrate efficient current-induced partial SOT magnetization switching in an epitaxial single layer of the oxide Weyl ferromagnet SRO. The inhomogeneous distribution of the spontaneous oxygen octahedral rotations, which are unique to complex oxides, leads to pronounced SOTs in a single SRO ferromagnetic layer. We obtain a small critical switching current density of $\sim 3.1 \times 10^6$ A cm$^{-2}$. The band inversion appears around the small gaps generated at band crossings near the SRO/STO interface due to the broken sublattice symmetry by the octahedral rotations. The inherent SOC in SRO accompanied by this phenomenon results in strong spin Berry curvature and, thus, a large intrinsic SHC due to the generation of *hot spots* of spin Berry curvature. Our findings highlight the immense potential for achieving giant SOTs through precise atom positioning in single crystals, unlocking a crucial pathway toward efficient and functional material systems for spin-orbitronics applications.

## 4. Experimental Section

*Sample Preparation*: We grew a 26 nm-thick SRO film on an STO (001) substrate using a custom-designed MBE setup equipped with multiple e-beam evaporators for Sr and Ru[34]. For growth, we precisely controlled the elemental fluxes by monitoring the flux rates with an electron-impact-emission-spectroscopy sensor and feeding the results back to the power supplies for the e-beam evaporators. Oxidation during growth was carried out using a mixture of oxygen (85%) and ozone (15%) gases; these gases were introduced through an alumina nozzle pointed at the substrate. Further information on the MBE setup and preparation of the substrate is described elsewhere. The clear Laue fringes obtained from X-ray diffraction (Figure S1, Supporting Information) and HAADF-STEM images shown in Figure 5a indicate the high crystallinity, a large coherent volume of the SRO film, and an abrupt interface between the SRO film and the STO substrate.

*Device preparation and electrical measurements*: We patterned the SRO/STO sample into a crossbar device with a channel width and length of 10 μm and 40 μm, respectively, via photolithography and argon ion milling. Afterward, Ag was sputtered to produce four electrode terminals, which also functioned as heat sinks. From the kink observed in the temperature dependence of the resistivity $\rho_{xx}$, the Curie temperature of the device was estimated to be ~150 K. For the SOT magnetization switching measurements, a Keithley 6221A was used as a pulsed current source.

*Theoretical calculations based on a tight-binding model*: We constructed a tight-binding model for SRO, following the schemes in refs. [29,42]. We consider the band structure of SRO near the SRO/STO interface, where half of the $t_{2g}$ band is filled because of the charge transfer of one electron from Ru to Ti near the SRO/STO interface[43]. Here, $Ru^{4+}$ in SRO has four electrons in the *d* orbital, whereas $Ti^{4+}$ in STO has none; these configurations are energetically favorable for the transfer of an electron from SRO to STO. This tight-binding model effectively reproduces the low-energy band structure, including the Weyl point structure under spontaneous magnetization. We calculated the band structure of SRO considering the nearest neighbor (NN) and next-nearest neighbor (NNN) direct Ru-Ru hopping, where the states with effective total angular momentum $J_{\text{eff}}$ values of 1/2 and 3/2 are mixed. Our tight-binding Hamiltonian $\mathcal{H}_k$ consists of the following four terms:

$$\mathcal{H}_k = \mathcal{H}_k^{\text{NN}} + \mathcal{H}_k^{\text{NNN}} + \mathcal{H}_k^{\text{SO}} + \mathcal{H}_k^{\text{exc}}. \quad (3)$$

Here, $\mathcal{H}_k^{\text{NN}}$ and $\mathcal{H}_k^{\text{NNN}}$ represent the hopping between the NN and NNN sites, respectively. $\mathcal{H}_k^{\text{SO}}$ denotes the spin–orbit coupling at each site. $\mathcal{H}_k^{\text{exc}}$ represents the effect of spin splitting by spontaneous magnetization.

We define *α* and *γ* as the angles of rotation of a $RuO_6$ octahedron around the *x*-axis and around the *z*-axis, respectively. Because the neighboring octahedrons located in the same *xz* plane rotate in opposite directions at the same angle, we consider a unit cell composed of four sublattices (1, 2, 3, and 4) of Ru sites, as shown in Figure 5a. Note that, for *α* = 0°, we first consider the pseudo-cubic unit cell of SRO, whose lattice parameters are $a' = b' = c' = 3.93$ nm[46]; these values are different from those of the orthorhombic unit cell (in the case of *α* > 0°).

As the basis for the Hamiltonian, we utilized three $t_{2g}$ orbitals $(d_{yz}, d_{zx}, d_{xy})$. Due to octahedral rotations, the crystal field is rotated in each sublattice, and thus, the directions of the $t_{2g}$ orbitals are different among the sublattices. Thus, to introduce octahedral rotations into the model, we consider the local coordinate $(x', y', z')$ for each sublattice (see details in Supporting Text 3, Supporting Information).

The spin Berry curvature $\Omega_{nk}^{y,s_y}$ arises around the band inversion points, where the band spacing $(\epsilon_{nk} - \epsilon_{mk})$ becomes small and the Bloch wave function sharply varies in the *k*- space. In the same manner, the dimensionless total spin Berry curvature $\Omega_{\boldsymbol{k}}^{y,s_y}$ is mapped to Figure S8 in the Supporting Information and shown in the range from –1 to +1 after scaling by applying the factor $2/\hbar a'c'$ to the original value.

**Supporting Information**

Supporting Information is available from the Wiley Online Library or from the author.

**Acknowledgments**

This work is supported by Grants-in-Aid for Scientific Research (No. 22H04948, 20H05650, 23H03802, 24H00409, 22K03538, 23K17882), JST CREST (JPMJCR1777), JST ERATO (JPMJER2202), and the Spintronics Research Network of Japan (Spin-RNJ).

**Conflict of Interest**

The authors declare no conflict of interest.

**Author contributions**

Sample preparation: H.H. and Y.K.W.; sample growth: Y.K.W., Y.T., H.Y., and Y.K.; measurements: H.H.; data analysis: H.H. and S.K-T.; calculation: Y.A.; theoretical modeling: Y.A., J.I., and M.Y.; writing and project planning: H.H., Y.A., Y.K.W., M.T., and S.O.

**Data Availability Statement**

The data that support the findings of this study are available from the corresponding author upon responsible request.

**Keywords**

spin Berry curvature, spintronics, spin-orbitronics, molecular beam epitaxy, oxide electronics, Weyl ferromagnet

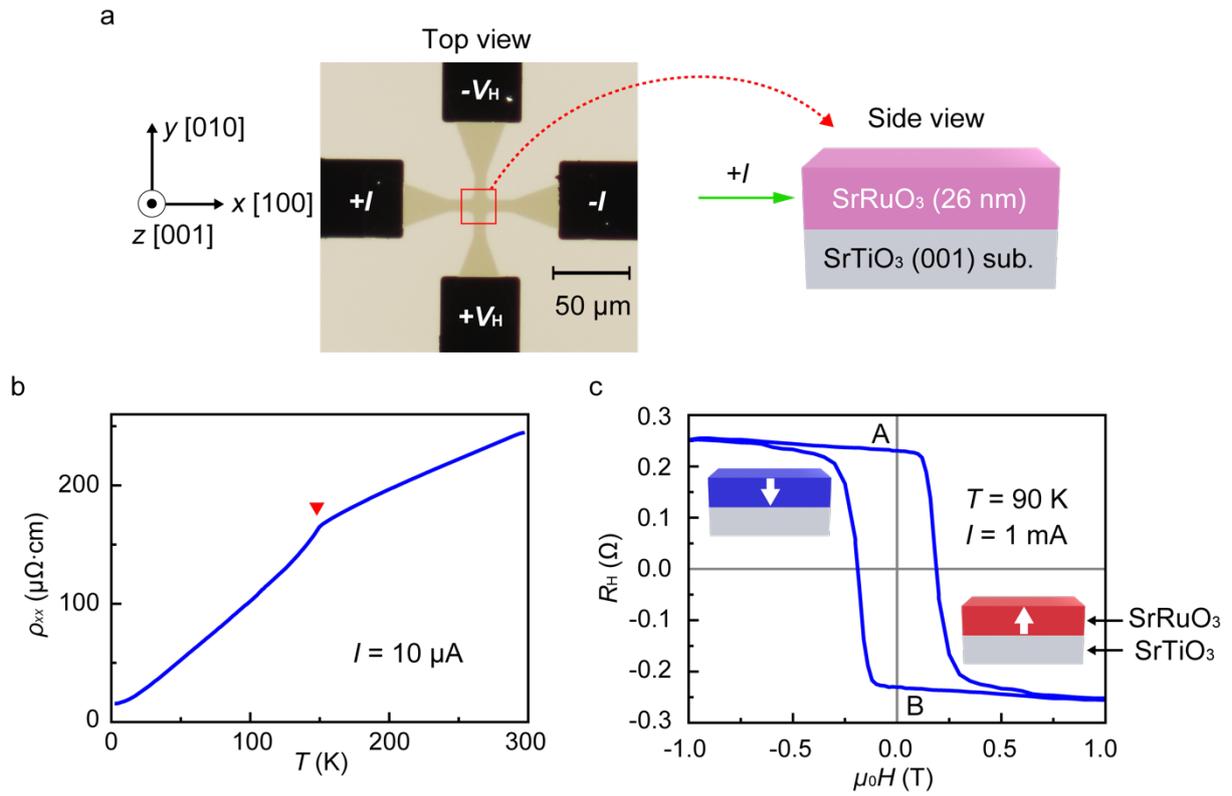

**Figure 1.** a) Optical microscope image of a crossbar device of SRO (26-nm thick) with Ag electrodes (black square pads). A current $+I$ is applied along the x-axis (//[100] of the STO substrate), and the Hall voltage between the terminals $+V_H$ and $-V_H$ is measured to derive the Hall resistance $R_H$ in magnetotransport measurements. b) Temperature ($T$) dependence of the resistivity ($\rho_{xx}$). The red triangle indicates the Curie temperature $T_C$ of ~ 150 K. c) $R_H$ vs. external magnetic field $\mu_0 H$ applied along the z-axis at 90 K, where $R_H$ is *negatively* proportional to $M_z$. The white arrows represent the magnetization directions at positions A and B.

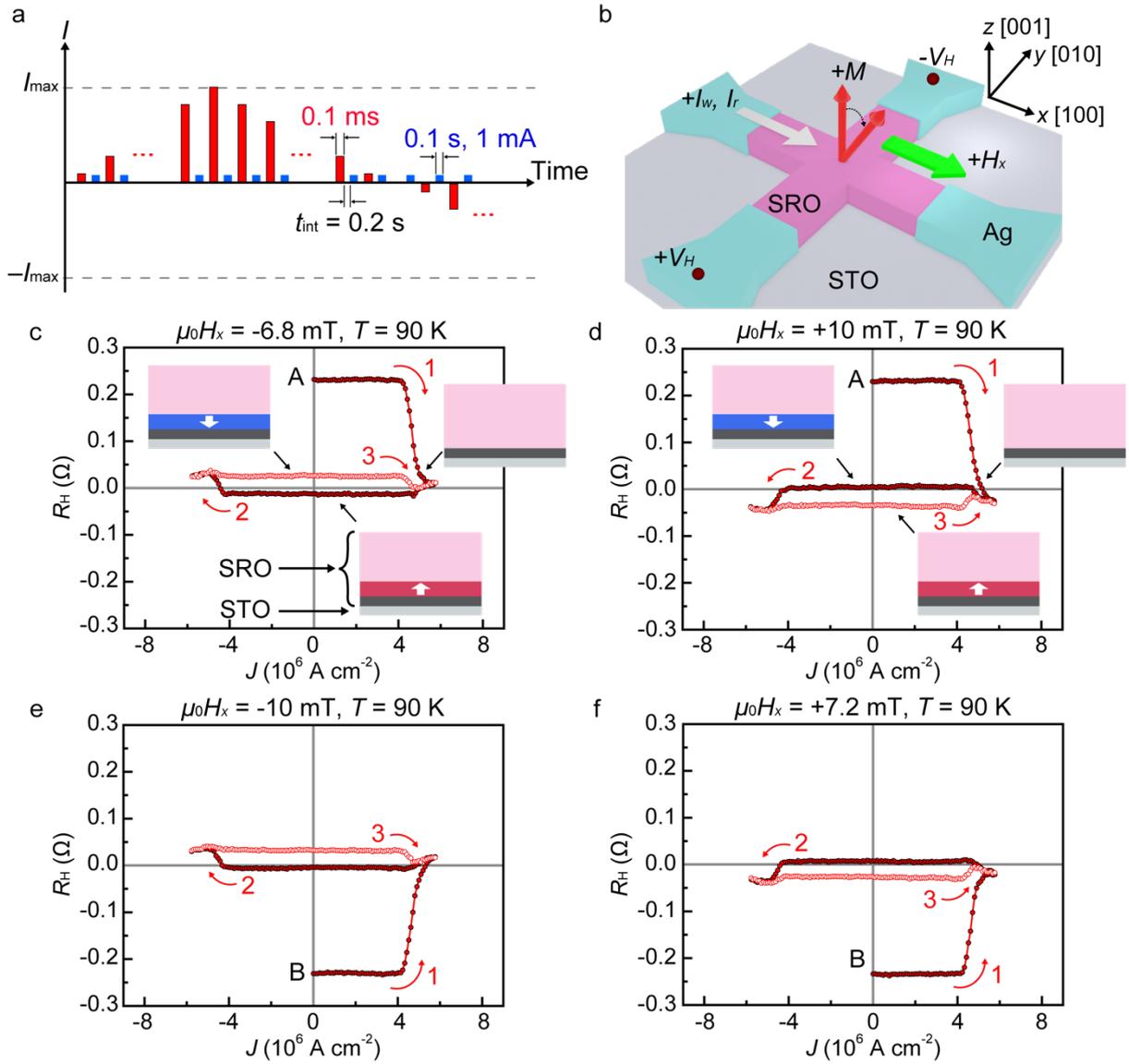

**Figure 2.** a) Sequence of the SOT magnetization switching measurements. b) Measurement configuration of SOT magnetization switching. The initial magnetization slightly tilts in the $x$ direction due to the supporting field $H_x$. c–f) $R_H$–$J$ loops obtained at $T$ = 90 K. Inserted schematic illustrations are the side view of schematic magnetization alignment in the SRO/STO (gray) heterostructure corresponding to the indicated points in processes 1, 2, and 3. The red and blue regions denote magnetic domains with the upward and downward magnetization directions, respectively. These regions correspond to ~8% of the 26 nm-thick SRO film near the SRO/STO interface, where the magnetization is likely switched in processes 2 and 3 at 90 K. The pink region is the mixed region of the upward and downward magnetization domains. The dark gray region represents the dead layer of the SRO film.

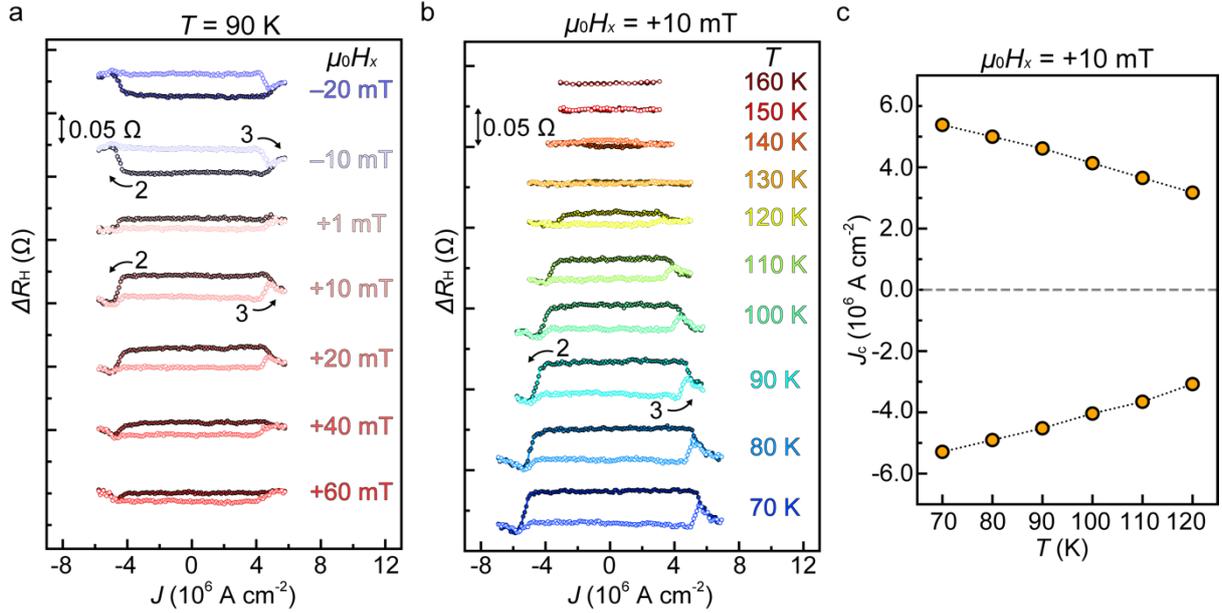

**Figure 3.** a) $\Delta R_H$–$J$ loops under various $H_x$ values at 90 K. b) $\varDelta R_H$–$J$ loops at various $T$ under $\mu_0 H_x = +10$ mT. In (a) and (b), the dark-colored (pale-colored) line in each loop corresponds to process 2 (3), as shown in Figure 2c–f. The arrows express the sweep directions. Before each measurement, the magnetization is initialized by applying a strong external magnetic field of 1 T along the −z direction, which corresponds to point A in Figure 2c,d. Here, process 1 is not shown, and only processes 2 and 3 are shown. The $\varDelta R_H$ obtained by subtracting the average value of the $R_H$ of each loop from the $R_H$ is plotted. c) Switching current density $J_c$ as a function of $T$ under a $\mu_0 H_x$ of +10 mT.

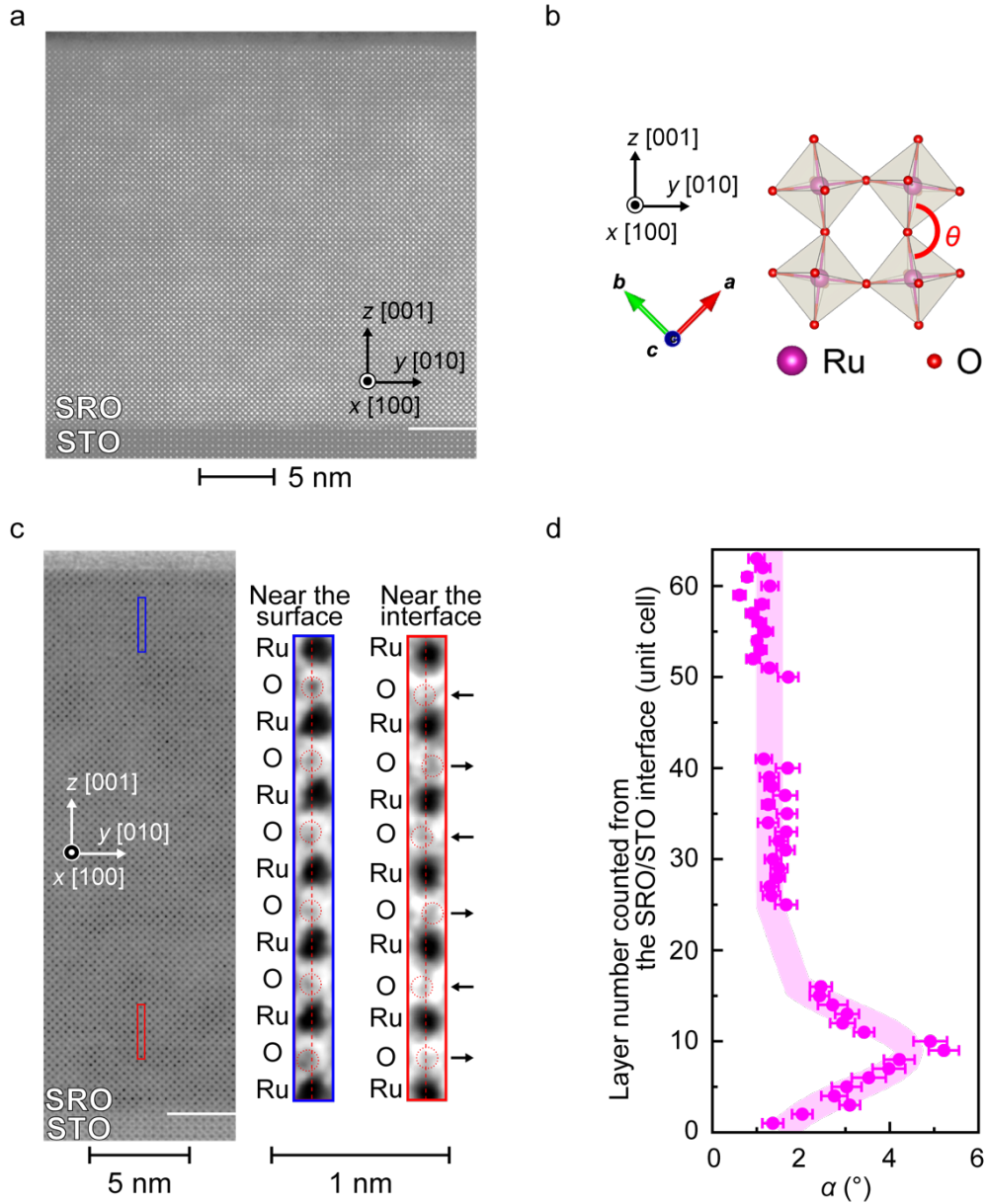

**Figure 4.** a) High-angle annular dark-field (HAADF-) STEM image of the SRO/STO (001) heterostructure. b) Lattice structure of RuO$_6$ octahedrons[47]. The coordinates are the same as those in Figure 2b. The arrows colored red, light green, and blue are the crystal axes of SRO on the STO (001) substrate. c) ABF-STEM image of the SRO/STO (001) heterostructure. The right images are magnified views of the area within the blue and red frames in the main image. Near the SRO/STO interface, the O atoms are displaced alternately along the ±$y$ direction. This provides evidence that oxygen octahedral rotation occurs. d) Distribution of the octahedral rotation angle $\alpha$ along the direction perpendicular to the film. The 0-th layer is defined as the oxygen atoms at the SRO/STO interface. The thick light pink line behind is a guide for the eyes.

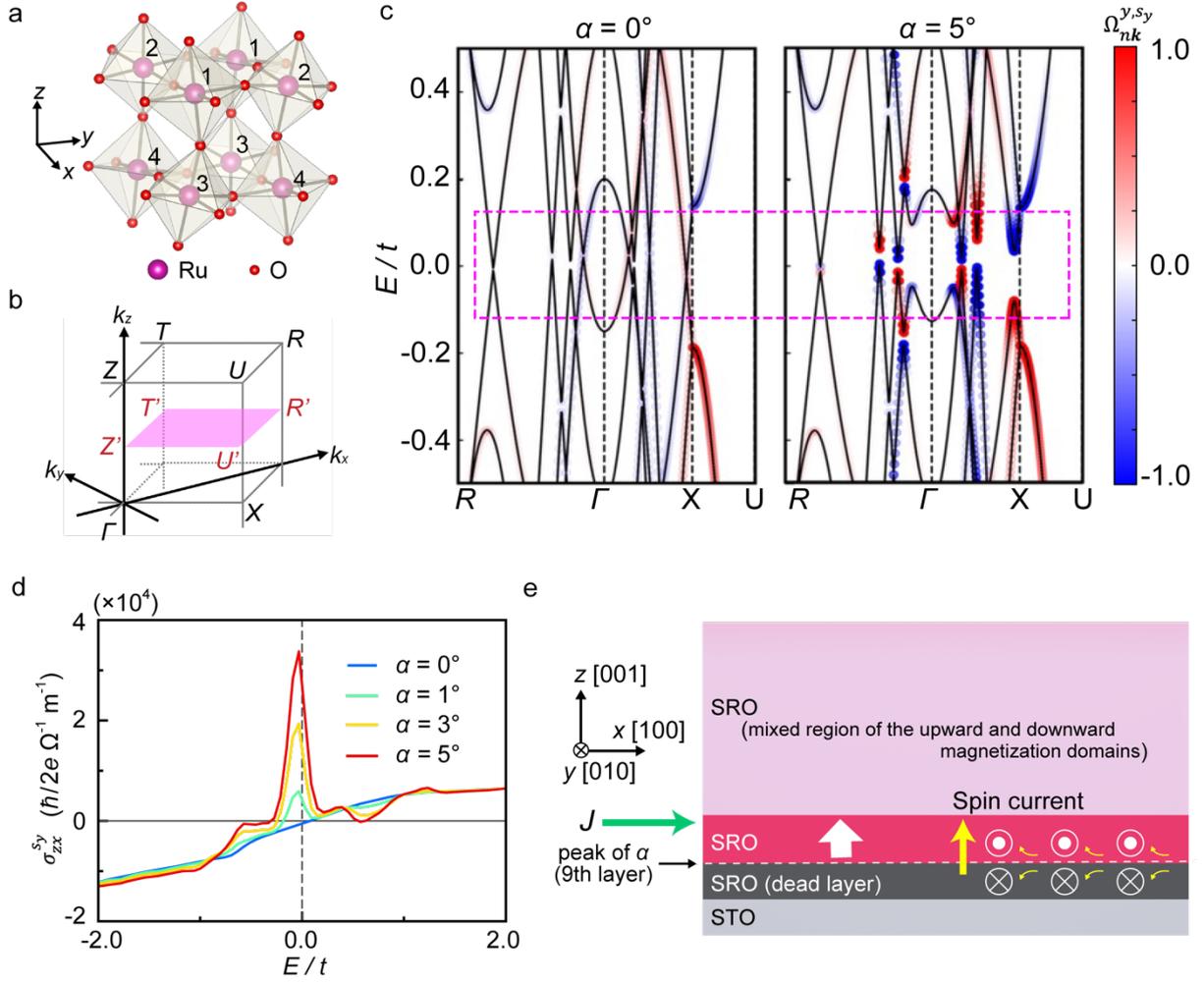

**Figure 5.** a) Illustration of the sublattices. b) Definition of the symmetric points in the ***k*** space. Each highly symmetric point corresponds to the Brillouin zone of the orthorhombic unit cell of SRO. ($k_x$, $k_y$, $k_z$) is defined from the crystal axis of the pseudo-cubic unit cell of SRO (*a'*, *b'*, *c'*). c) Band structure and spin Berry curvature $\Omega_{n\bm{k}}^{y,s_y}$ of SRO when the octahedral rotation angle $\alpha = 0°$ and $\alpha = 5°$. The value of $\Omega_{n\bm{k}}^{y,s_y}$ is a dimensionless quantity scaled by the factor $2/\hbar a'c'$ to the original value, has the dimension of a product of angular momentum and length squared $\hbar \cdot m^2$, is defined by the second derivative in ***k***-space, and is shown in the range from –1 to +1, expressed as the color of the dots. d) Intrinsic contribution of spin Hall conductivity (SHC) for various $\alpha$. e) Illustration of the spin Hall effect and induced magnetization switching near the SRO/STO interface. The red region represents the upward magnetization region, where the SOT switching occurs. Here, the spin directions along the *y*-axis for the positive writing current are illustrated with white marks. The pale pink region is the mixed region of the upward and downward magnetization domains.

**Supporting Information**

**Single-layer spin-orbit-torque magnetization switching due to spin Berry curvature generated by minute spontaneous atomic displacement in a Weyl oxide**

*Hiroto Horiuchi, Yasufumi Araki, Yuki K. Wakabayashi, Jun'ichi Ieda,*
*Michihiko Yamanouchi, Shingo Kaneta-Takada, Yoshitaka Taniyasu, Hideki Yamamoto,*
*Yoshiharu Krockenberger, Masaaki Tanaka, and Shinobu Ohya*

**Supporting Text 1: Sample characterizations**

The out-of-plane X-ray diffraction (XRD) pattern for the $SrRuO_3$ (SRO)/$SrTiO_3$ (STO) heterostructure shows peaks with Laue (Kiessig) fringes (see Figure S1), confirming that the sample is single phase with an abrupt interface.

The temperature dependence of magnetic moment obtained by a superconducting quantum interface device (SQUID) shows that the Curie temperature ($T_C$) of the SRO film is around 150 K (see Figure S2), which agrees with the temperature dependence of the resistivity shown in Figure 1b.

The SRO film has perpendicular magnetic anisotropy (PMA), as shown by the rectangular hysteresis loops of the anomalous Hall effect (AHE) (see Figure S3). The Hall resistance $R_H$ is *negatively* proportional to the perpendicular component of magnetization in the temperature range from 3.7 K to 120 K.

**Supporting Text 2: SOT-magnetization switching measurements**

Before each SOT-magnetization switching measurement, we applied a large external magnetic field $\mu_0 H$ of 1 T along the +z or –z direction to align the magnetization in those directions as an initialization process. As shown in Figure S4a,c,e,g, each measurement starts from the saturated $R_H$ values with the magnetization directions along the –z (point A) and +z (point B) directions.

After one measurement sequence (processes 1, 2, and 3) was completed, we repeated the same processes 1, 2, and 3 under the same in-plane magnetic field $H_x$ without initialization. Here, each measurement started with the $R_H$ value obtained at the end of the previous sequence. In those repeated sequences, we observed the same hysteresis loop as that obtained in the first sequence (Figure S4b,d,f,h). This result indicates that the magnetization is stably switched and that the switching process is non-volatile.

As shown in Figure 2c–f, the center of the $R_H$ – current density ($J$) loops deviates from $R_H = 0$ Ω depending on the sign and magnitude of $H_x$. This offset is due to a slight misalignment of the sample, whose surface is slightly displaced in the $xz$ plane by angle $\varphi$ from the direction of $H_x$, as shown in Figure S5a. In fact, with increasing $\varphi$ in the positive direction from 0°, the offset shifts in the plus direction of $R_H$ for $H_x < 0$ (Figure S5b) and in the minus direction for $H_x > 0$ (Figure S5c). This phenomenon originates from the multidomain region shown as pale pink in Figure 2c,d.

To understand this behavior, we consider a multidomain structure where +z- and –z-oriented magnetization domains are located alternatively, as reported for SRO[S1]. As shown in the lower inset of Figure S5c, when $H_x > 0$ and $\varphi > 0$, the magnetization oriented along the –z direction is more largely tilted towards the $x$ direction than that oriented along the +z direction because the +z direction is closer to the $H_x$ direction than the –z direction. Here, with increasing $\varphi$, the magnetization direction changes from the light-colored white arrow to the dark-colored white arrow in the downward magnetization domain. Under this configuration, the magnetization oriented along the +z direction is not significantly affected by the increase in $\varphi$ (see the upward magnetization domain of the bottom inset in Figure S5c). Thus, the total magnetization of the multidomain region becomes positive. As a result, due to the negative AHE coefficient, the hysteresis loop has a negative offset when $H_x > 0$ and $\varphi > 0$ (Figure S5c). From this measurement, $\varphi$ is estimated to be 6°. Here, as clearly seen in Figure S5b,c, the shape of the hysteresis loops does not change even though the offset changes. Hence, this misalignment of the sample does not affect the switching process itself.

**Supporting Text 3: Theoretical calculation of the spin Berry curvature and spin Hall conductivity in SRO**

We theoretically calculated the spin Berry curvature and the spin Hall conductivity (SHC) based on the scheme in refs. [S2,S3]. We considered the band structure of SRO near the SRO/STO interface, where half of the $t_{2g}$ band is filled due to the charge transfer of one electron from Ru to Ti near the SRO/STO interface[S4].

As the basis for the Hamiltonian shown in the Experimental Section, we utilized the three $t_{2g}$ orbitals $(d_{yz}, d_{zx}, d_{xy})$. Due to octahedral rotations, the crystal field direction and, thus, the directions of the $t_{2g}$ orbitals are different among the sublattices. Thus, we considered the local coordinate $(x', y', z')$ for each sublattice. It is related to the original coordinate $(x, y, z)$, which is equivalent to the crystal axes of the pseudocubic SRO cell, (see Figure 5a) as

$$\begin{pmatrix} x' \\ y' \\ z' \end{pmatrix} = \begin{pmatrix} \cos\gamma_i & -\sin\gamma_i & 0 \\ \cos\alpha_i \sin\gamma_i & \cos\alpha_i \cos\gamma_i & \sin\alpha_i \\ -\sin\alpha_i \sin\gamma_i & -\sin\alpha_i \cos\gamma_i & \cos\alpha_i \end{pmatrix} \begin{pmatrix} x \\ y \\ z \end{pmatrix}, \quad (S1)$$

where $\alpha_i$ and $\gamma_i$ are the rotation angles around the $x$ and the $z$ axes, respectively, in sublattice $i$ (= 1,…,4, see Figure 5a). $\alpha_i$ and $\gamma_i$ satisfy the following conditions;

$$\alpha_1 = -\alpha_2 = -\alpha_3 = \alpha_4 \equiv \alpha, \quad \gamma_1 = -\gamma_2 = \gamma_3 = -\gamma_4 \equiv \gamma. \quad (S2)$$

In this coordinate, the $t_{2g}$ orbitals $(d_{y'z'}, d_{z'x'}, d_{x'y'})$ are given as the linear combinations of $d_{yz}, d_{zx}, d_{xy}, d_{x^2-y^2}$, and $d_{3r^2-z^2}$ defined in the original coordinate as

$$\begin{pmatrix} d_{y'z'} \\ d_{z'x'} \\ d_{x'y'} \end{pmatrix}$$

$$= \begin{pmatrix} \cos 2\alpha_i \cos\gamma_i & \cos 2\alpha_i \sin\gamma_i & -\frac{1}{2}\sin 2\alpha_i \sin 2\gamma_i & \frac{1}{2}\sin 2\alpha_i \cos 2\gamma_i & \frac{\sqrt{3}}{2}\sin 2\alpha_i \\ -\cos\alpha_i \sin\gamma_i & \cos\alpha_i \cos\gamma_i & -\sin\alpha_i \cos 2\gamma_i & -\sin\alpha_i \sin 2\gamma_i & 0 \\ -\sin\alpha_i \sin\gamma_i & \sin\alpha_i \cos\gamma_i & \cos\alpha_i \cos 2\gamma_i & \cos\alpha_i \sin 2\gamma_i & 0 \end{pmatrix} \begin{pmatrix} d_{yz} \\ d_{zx} \\ d_{xy} \\ d_{x^2-y^2} \\ d_{3r^2-z^2} \end{pmatrix}$$

(S3)

We denote this 3×5 matrix as $C_{lm}^i$ (with $l = y'z', z'x', x'y'$ and $m = yz, zx, xy, x^2 - y^2, 3r^2 - z^2$), where the superscript $i$ denotes the sublattice site index. We used this coordinate transformation to construct the tight-binding model with octahedral rotation.

The hopping terms are constructed using Slater-Koster's method. When we consider the hopping between orbital $d_l^i$ on site $i$ and orbital $d_m^j$ on site $j$ (with $l, m = yz, zx, xy, x^2 - y^2, 3r^2$

– $z^2$), the hopping amplitude $t_{lm}^{ij}$ is given by the linear combination of those of the σ, π, and δ-bondings, $t_\sigma^{ij}$, $t_\pi^{ij}$, and $t_\delta^{ij}$, respectively. The structure of $t_{lm}^{ij}$ depends on the orientation of site $i$ from site $j$. The hopping amplitude between the $t_{2g}$ orbitals in the local coordinate ($\tilde{t}_{lm}^{ij}$) becomes

$$\tilde{t}_{lm}^{ij} = \sum_{o,p} C_{lo}^i \, t_{op}^{ij} \, {}^t C_{pm}^j \,, \tag{S4}$$

where, the letter $t$ on the left shoulder of $C$ means the transposed matrix of $C$, and $o$ and $p$ take $yz$, $zx$, $xy$, $x^2 - y^2$, and $3r^2 - z^2$. We used $\tilde{t}_{lm}^{ij}$ for the NN and NNN hopping terms.

Here, neighboring octahedrons located in the same $yz$ plane rotate in opposite directions with the same angle. Hence, we considered a unit cell consisting of four sublattices 1–4 shown in Figure 5a. Since the volume of the unit cell quadruples, the Brillouin zone is folded into a quarter of that for a single octahedron unit cell, by which band crossings appear at and around high-symmetric points, for example, at $T$, $U$, $X$, $Z$ and around $\Gamma$ (see red and blue dashed lines in Figure S6).

In general, octahedrons also rotate around the $z$ axis in addition to the $x$ axis. As seen in Figure S7, increasing only the $z$-axis rotation, whose angle is defined as $\gamma$, does not significantly enhance the SHC $\sigma_{zx}^{s_y}$ in comparison with the case when increasing only the $x$-axis rotation angle $\alpha$.

The octahedral rotations enhance the total spin Berry curvature, intensifying the spin Hall effect. Figure S8 shows the Fermi contours in the **k** plane of $Z'U'R'T'$ and the sum of spin Berry curvature $\Omega_{n\mathbf{k}}^{y,s_y}$ (color scale) defined as $\Omega_{\mathbf{k}}^{y,s_y} = \sum_n f(\epsilon_{n\mathbf{k}}) \Omega_{n\mathbf{k}}^{y,s_y}$, where $f(\epsilon_{n\mathbf{k}})$ is the Fermi distribution function. Temperature $T$ is set at 0 K. Fermi contours named A that can be seen when $\alpha = 0°$ (bottom left area surrounded by the green broken circle in Figure S8b) are not visible when $\alpha = 5°$ (Figure S8c), indicating that a small gap opens for this band. Fermi contours named B that are very close to each other when $\alpha = 0°$ (upper right area surrounded by the green broken circle in Figure S8b) merge into one when $\alpha = 5°$ (Figure S8c), indicating the splitting bands that cross slightly below the $E_F$ are lifted up due to gap opening by octahedral rotations. By comparing Figure S8b ($\alpha = 0°$) and Figure S8c ($\alpha = 5°$), one can see that $\Omega_{\mathbf{k}}^{y,s_y}$ is strongly enhanced around the Fermi contours by octahedral rotations.

Changes above in the electronic structure and the enhancement of $\Omega_{\mathbf{k}}^{y,s_y}$ originate from the band repulsion and hybridization because of the broken sub-lattice symmetry by octahedral rotations. Here, we discuss the mechanism of how the octahedral rotation influences the emergence of spin Berry curvature by considering the symmetry of the system. We focus on the band crossing and repulsion structure around $E = 0$.

In the absence of octahedral rotations, we find many Fermi contours that are almost doubly degenerate around $E = 0$. Such a structure around $E = 0$ can be understood from the "approximate" chiral symmetry and the sublattice symmetry. The dominant parts $\mathcal{H}_k^{NN}$ and $\mathcal{H}_k^{exc}$ in the tight-binding Hamiltonian are antisymmetric under the hypothetically defined unitary transformation $\Gamma = Q s_y$,

$$\Gamma \mathcal{H}_k^{NN} \Gamma^{-1} = -\mathcal{H}_k^{NN}, \qquad \Gamma \mathcal{H}_k^{exc} \Gamma^{-1} = -\mathcal{H}_k^{exc}, \tag{S5}$$

where $Q$ multiplies the phase factor $+1$ or $-1$ on each sublattice,

$$Q: (d_1, d_2, d_3, d_4) \to (d_1, -d_2, -d_3, d_4). \tag{S6}$$

Equation (S5) means that $\Gamma$ serves as the chiral symmetry for $\mathcal{H}_k^{NN}$ and $\mathcal{H}_k^{exc}$. If we consider only $\mathcal{H}_k^{NN}$ and $\mathcal{H}_k^{exc}$, we can rigorously conclude from the chiral symmetry that the states at $E = 0$ are doubly degenerate. Once we introduce the terms $\mathcal{H}_k^{NNN}$ and $\mathcal{H}_k^{SO}$, they slightly violate the chiral symmetry. Nevertheless, since the system still satisfies the sublattice symmetries, which are defined by the half translation of the unit cell,

$$T_y^{1/2}: (d_1, d_2, d_3, d_4) \to (d_2, d_1, d_4, d_3) \tag{S7}$$

$$T_z^{1/2}: (d_1, d_2, d_3, d_4) \to (d_3, d_4, d_1, d_2). \tag{S8}$$

Therefore, the double degeneracies mentioned above are protected by the sublattice symmetries, while their energies are slightly lifted from $E = 0$. We plot the eigenvalues of $T_y^{1/2}$ for each band around $E = 0$, as shown in Figure S9b. We find several bands with $T_y^{1/2} = +1$ and $T_y^{1/2} = -1$ crossing around $E = 0$, which are not gapped out due to the symmetry $T_y^{1/2}$. These crossing bands are almost spin polarized, with $s_z = \uparrow$ and $s_z = \downarrow$, as shown in Figure S9c.

Now we introduce the effect of the octahedral rotations. Since the sublattice symmetry $T_y^{1/2}$ (or $T_z^{1/2}$) protecting the band crossings is broken, the bands with $T_y^{1/2} = +$ and $T_y^{1/2} = -$ are now hybridized and gapped out, as we show in Figure S9e. As a consequence, the bands with $s_z = \uparrow$ and $s_z = \downarrow$ are hybridized, and thus the direction of spins on each band is drastically altered around the hybridization points in momentum space, as shown in Figure S9f. In other words, we can regard that the effect of spin-orbit coupling is magnified drastically around the hybridization points, which forms *hot spots* of spin Berry curvature.

**Supporting Text 4: Estimation of the SHC**

The ratio of $\sigma_{zx}^{s_y}$ to the longitudinal conductivity $\sigma_{xx}$ is defined as the spin Hall angle $\theta_{SH}$, which is estimated as

$$\theta_{SH} = \frac{2e}{\hbar} \cdot M_s t_{FM} \cdot \frac{H_c}{J_c}, \qquad (S9)$$

where $e$, $\hbar$, $M_s$, $t_{FM}$, $H_c$, and $J_c$ are the elementary charge, reduced plank constant, saturated magnetization, thickness of the switched area in the SRO film (see Figure 5e), coercive field, and the critical switching current, respectively[S5,S6]. Equation (S9) refers to the efficiency of SOT-magnetization switching via local domain wall depinning. By substituting $M_s = 1.42 \times 10^5$ A m$^{-1}$, $t_{FM} = 2.08$ nm (= ~ 8 % of the 26 nm-thick SRO film), $H_c = 1800$ Oe, and $J_c = 3.1 \times 10^6$ A cm$^{-2}$, all of which are obtained experimentally at $T = 90$ K, to equation (S1), we obtain $\theta_{SH}$ ~ 0.83. Thus, from the relation

$$\theta_{SH} = \left(\frac{2e}{\hbar}\right) \sigma_{zx}^{s_y} \cdot \sigma_{xx}, \qquad (S10)$$

we can roughly estimate $\sigma_{zx}^{s_y}$ ~ $9.0 \times 10^5$ ($\hbar/2e$) $\Omega^{-1}$ m$^{-1}$ at 90 K. Here, we neglected the effect of the spin diffusion at the interface between ferromagnet and non-magnet that is considered in bilayer systems and is determined by spin transparency[S7]. The estimated value of $\sigma_{zx}^{s_y}$ is about 3.9 times larger than that for a Co/SRO bilayer system[S7].

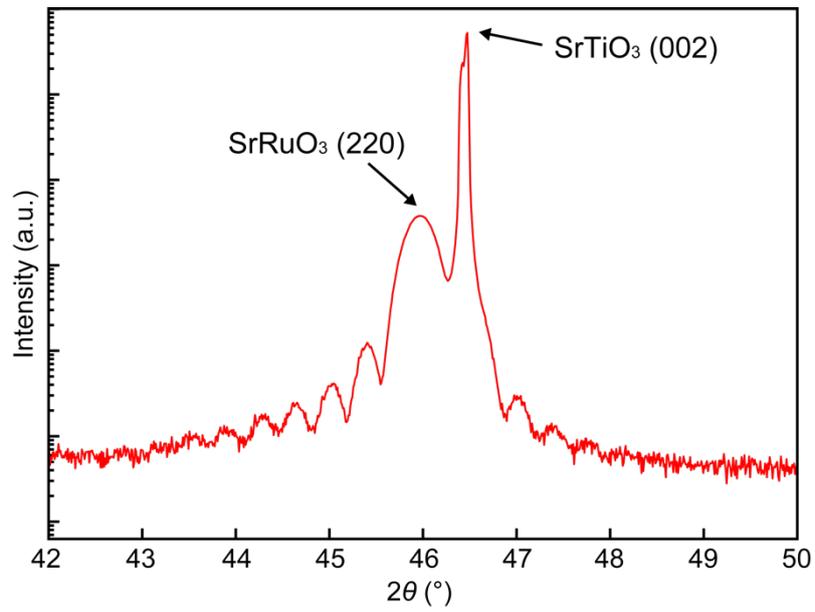

**Figure S1.** X-ray diffraction (XRD) patterns of SRO on the STO substrate. XRD $2\theta$–$\omega$ scan of SRO (26 nm)/STO is shown.

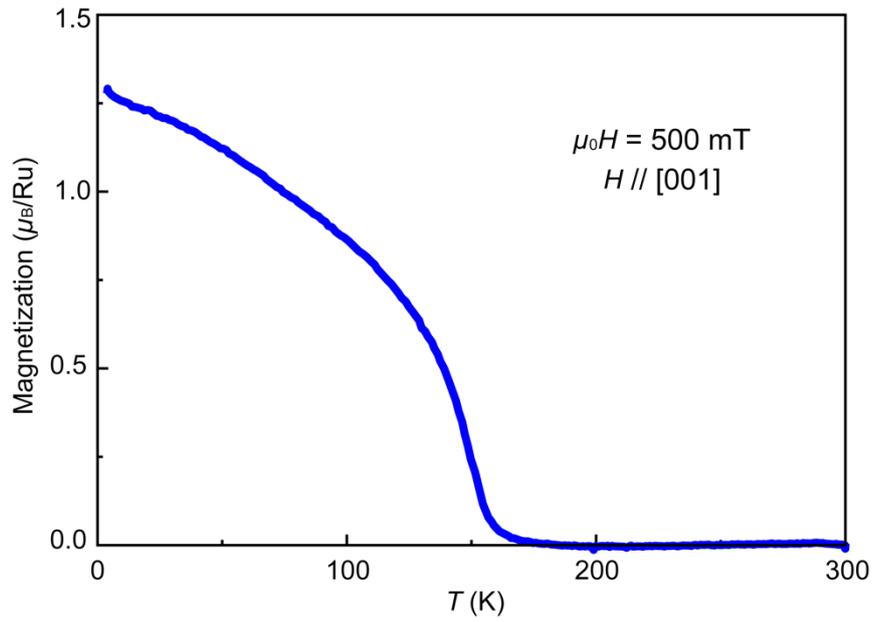

**Figure S2.** Temperature (*T*) dependence of the magnetization. The external magnetic field of 500 mT is applied along the *z* ([001] of the STO substrate) direction.

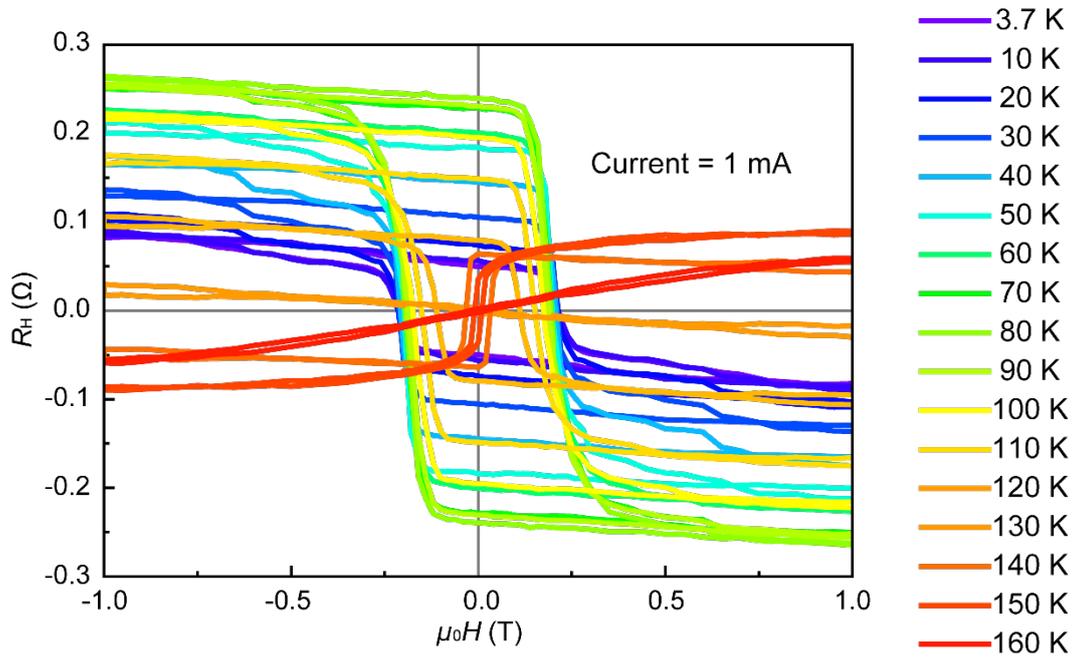

**Figure S3.** Temperature dependence of the Anomalous Hall effect. The magnetic field is applied along the $z$ direction. The Hall resistance $R_H$ is negatively proportional to the perpendicular component of the magnetization in the temperature range from 3.7 K to 120 K.

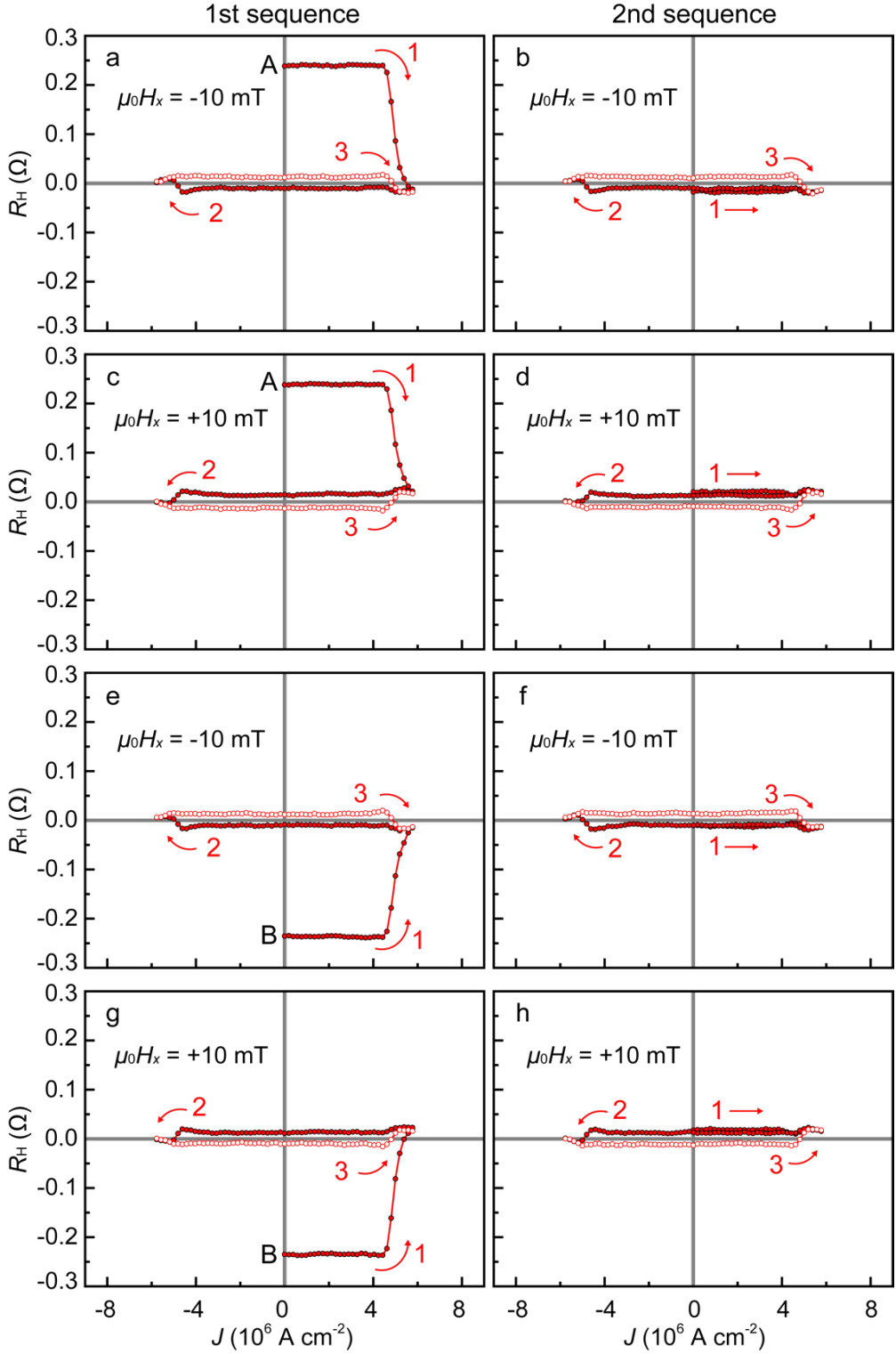

**Figure S4.** a),c),e),g) $R_H$–$J$ loops obtained after the initialization process. The values of $R_H$ at A and B correspond to the initial states where the magnetization is aligned along the –$z$ and +$z$ directions, respectively. The measurement process proceeds in the order of 1→2→3. b),d),f),h) $R_H$–$J$ loops obtained in the second sequence of processes 1, 2, and 3. The above data are taken for a different device that has the same SRO thickness and nearly the same magnetic properties as that shown in the main manuscript. All measurements were carried out at 90 K.

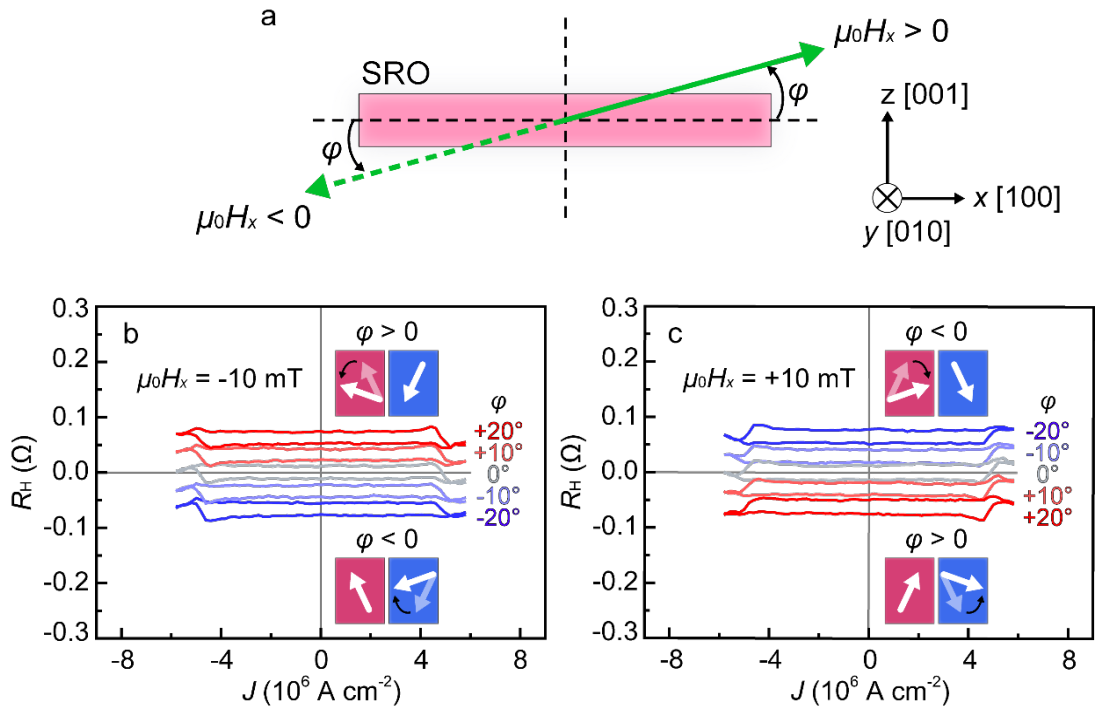

**Figure S5.** a) Illustration of the misalignment of the sample plane from the $H_x$ direction. The pink rectangle is the side view of the single-layer SRO film. b),c) $R_H$–$J$ loops obtained at different misalignment angles $\varphi$ for (b) $\mu_0 H_x = -10$ mT and (c) $\mu_0 H_x = +10$ mT. Insets are the side views of the expected direction of upward (red) and downward (blue) magnetization domains in the mixed region in Figure 5e for $\varphi > 0$ and $\varphi < 0$. The direction of magnetization within each domain transitions from the lighter-colored white arrow to the darker-colored white arrow as $|\varphi|$ increases.

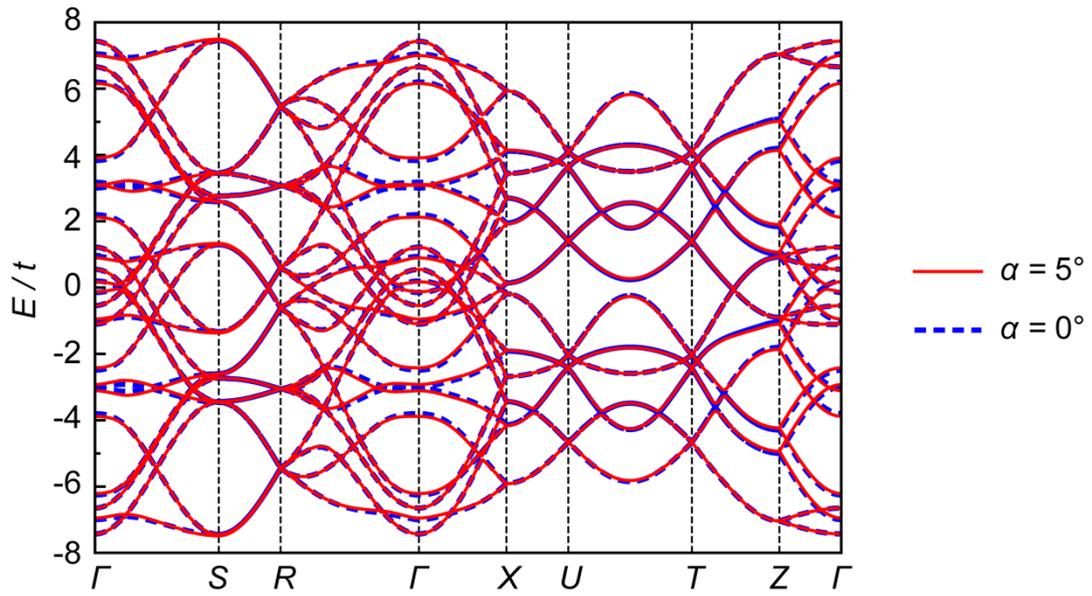

**Figure S6.** Band structure calculation results. Band structure of SRO for $\alpha = 0°$ and $\alpha = 5°$.

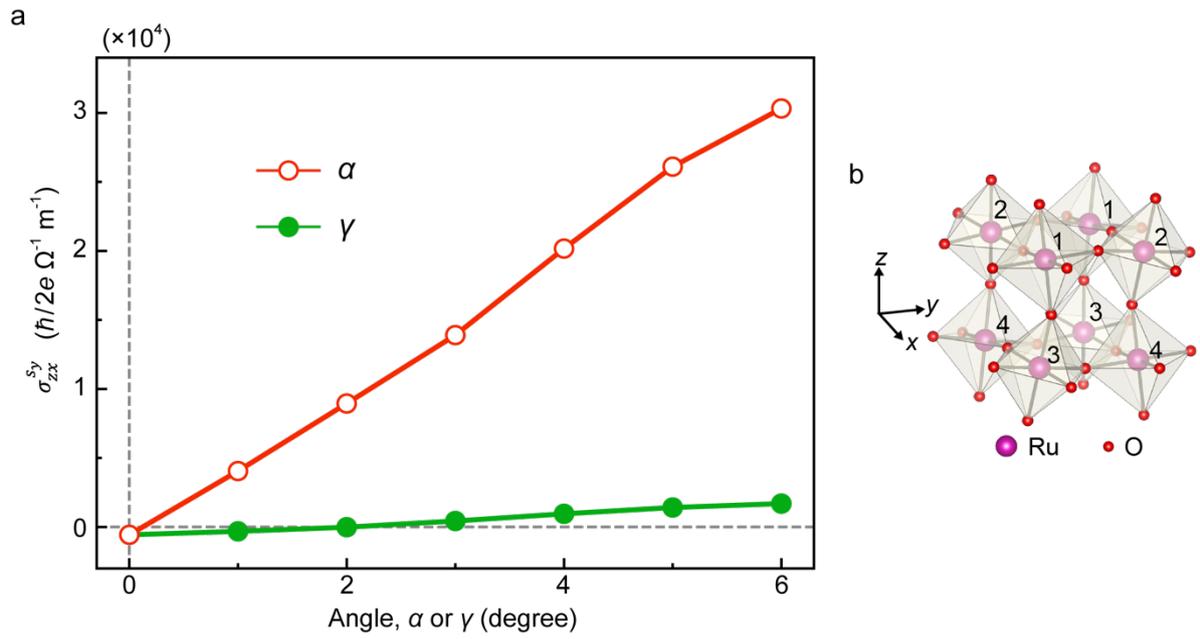

**Figure S7.** a) Change in $\sigma_{zx}^{s_y}$ of SRO when changing $\alpha$ for $\gamma = 0°$ (red) and when changing $\gamma$ for $\alpha = 0°$ (green) at $E/t = 0$. b) Illustration of the sublattices of SRO.

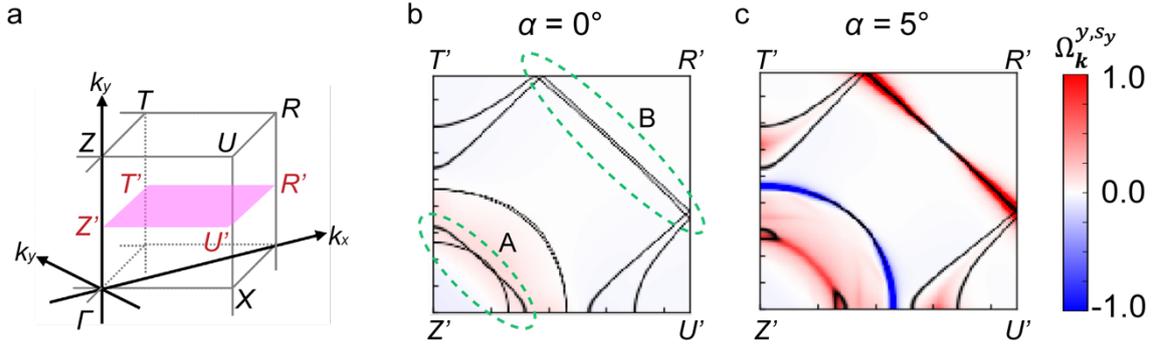

**Figure S8.** a) Definition of the symmetric points in the $\boldsymbol{k}$-space. b),c) Fermi contours and distribution of the spin Berry curvature $\Omega_{\boldsymbol{k}}^{y,s_y} = \sum_n f(\epsilon_n(\boldsymbol{k}))\Omega_{n\boldsymbol{k}}^{y,s_y}$ in the $\boldsymbol{k}$-space at the energy $E = 0$ for (b) $\alpha = 0°$ and (c) $\alpha = 5°$ calculated from the model. The value of $\Omega_{\boldsymbol{k}}^{y,s_y}$, which is a dimensionless quantity scaled using the lattice parameter (see Section 4), is shown in the range from –1 to +1, expressed as the color scale.

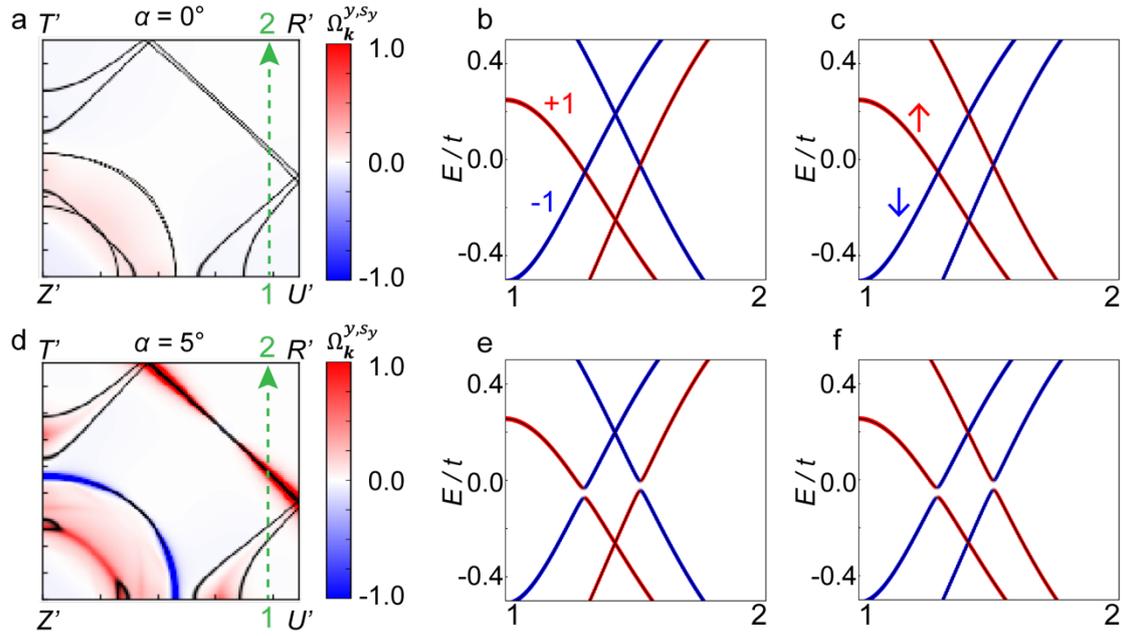

**Figure S9.** a),d) Fermi contours and distribution of $\Omega_{k}^{y,s_y} = \sum_n f(\epsilon_n(\boldsymbol{k}))\Omega_{n\boldsymbol{k}}^{y,s_y}$ in the $\boldsymbol{k}$-space at the energy $E = 0$ for (a) $\alpha = 0°$ and (d) $\alpha = 5°$ as shown in Figure S8. b),e) Distributions of the parity under sublattice transformation (half-unit cell translation) $T_y^{1/2}$ on each band for (b) $\alpha = 0°$ and (e) $\alpha = 5°$ along the green arrow in (a) and (d), respectively. +1 (red) and –1 (blue) indicates if the Bloch state is even or odd under $T_y^{1/2}$, respectively. c),f) Distributions of the spin polarization $s_z$ for (c) $\alpha = 0°$ and (f) $\alpha = 5°$ along the green arrow in (a) and (d), respectively. ↑ (red) and ↓ (blue) indicate up- and down-spin, respectively.

**SI References**